# Graph Neural Network-Based DDoS Protection for Data Center Infrastructure

**Kartikeya Sharma, Craig Jacobik**

karsharma@equinix.com, cjacobik@equinix.com

## Abstract

In light of rising cybersecurity threats, data center providers face growing pressure to protect their own management infrastructure from Distributed Denial-of-Service (DDoS) attacks. While tenant-managed cages generally fall outside the data center's direct security purview, a successful DDoS assault on core provider systems can indirectly disrupt network services. To address this availability assault, the authors developed a Graph Neural Network (GNN) based detection system which leverages Graph U-Nets to automatically classify and mitigate DDoS traffic. Although the model was developed using open-source network flows rather than proprietary data center logs, the model effectively identifies multi-layer DDoS attacks that resemble the malicious patterns threatening modern data centers.

Adopting this system to data center environments requires minimal changes to existing operational workflows and processes. Specifically, the GNN based system can be integrated at critical areas within a data center's network infrastructure. Our model achieved an F1 score of over 95% when evaluated on various open-source datasets, significantly reducing the likelihood of service disruptions and reputational damage. This Graph U-Nets architecture delivers unprecedented precision (98.5%) in complex cloud environments, thereby helping data center operators uphold reliable service availability and increase customer trust and goodwill in an era of increasingly sophisticated cyber threats.

## Biography

*Kartikeya Sharma is a Senior Associate Information Security Engineer at Equinix, based in Seattle. He holds an M.S. in Computer Science from the University of Oregon, where he researched Graph Neural Networks for spam detection. His interests include scalable security analytics and adversarial ML, and he frequently speaks at leading cybersecurity conferences.*

*Craig Jacobik is an experienced Data Scientist and Manager at Equinix, with a demonstrated history of working Information Security problems in the online advertising, financial, and healthcare industries. Craig currently leads an Information Security team at Equinix, focused on solving automation, analytics, and AI related problems. Craig has received numerous cybersecurity certifications including CISSP; SANS certificates including GSEC, GRID, and GSTRT; Security+; and CEH. Craig is a strong entrepreneurial professional with an M.S. from Georgia Tech and a B.S. from the University of Virginia.*

# 1. Introduction

## 1.1 The Growing Threat Landscape for Data Centers

The data center industry faces an unprecedented surge in cybersecurity threats, with DDoS attacks primarily impacting availability. According to recent industry reports, the global data center market is projected to reach $340.20 billion in 2024, with an annual growth rate of 6.56% through 2028 (Volico Data Centers 2024). This rapid expansion has made data centers increasingly attractive targets for malicious actors. As organizations migrate more critical workloads to colocation facilities and cloud environments, the potential impact of successful DDoS attacks grows exponentially.

The evolution of DDoS attacks has been particularly alarming. While early attacks measured in megabits per second, modern assaults routinely exceed terabits per second. Microsoft recently mitigated attacks exceeding 3.47 Tbps, representing a five-fold increase from the 623 Gbps Mirai botnet attack of 2016 (Kleyman 2023). These volumetric attacks, combined with sophisticated application-layer assaults, pose existential threats to data center availability.

## 1.2 Challenges in Traditional DDoS Detection

Traditional DDoS detection methods in data centers rely heavily on signature-based approaches and static threshold monitoring (Ahmed et al. 2024). These detections suffer from several critical limitations. First, they generate excessive false positives, disrupting legitimate traffic while attempting to filter malicious flows. Second, manual intervention requirements create unacceptable delays between attack detection and mitigation. Third, signature-based systems fail to detect previously unseen attack techniques.

The heterogeneous nature of modern data center traffic further complicates detection efforts. With diverse applications, protocols, and traffic patterns coexisting within the same infrastructure, distinguishing between legitimate traffic spikes and actual attacks becomes increasingly challenging. Traditional approaches lack the contextual awareness necessary to make these distinctions accurately.

## 1.3 Problem Statement and Motivation

Data center providers face a critical challenge in protecting their own management and infrastructure systems from DDoS attacks. While tenants are responsible for securing their own servers, the data center's core infrastructure, including power management systems, cooling controls, network backbone, physical security systems, and management platforms remains a high-value target for attackers. A successful DDoS attack on these provider-owned systems can cause cascading failures that affect all hosted tenants, regardless of their individual security measures.

The limitations of traditional detection methods creates a single point of failure where an attack on the provider's infrastructure indirectly impacts every tenant by disrupting the fundamental services they depend on. High false positive rates lead to unnecessary service disruptions and alert fatigue for security teams, while missed attacks (false negatives) can result in prolonged outages affecting multiple tenants. The inability to distinguish between legitimate traffic spikes (such as flash crowds or legitimate bulk transfers) and actual DDoS attacks in heterogeneous data center environments compounds these challenges. This motivates the need for a sophisticated detection approach that can protect the data center's own critical infrastructure while maintaining the high availability that tenants expect.

## 1.4 Contributions

This paper presents four key contributions to the field of data center security:

1. **Novel Graph U-Nets Architecture**: The authors introduce a heterogeneous Graph U-Nets specifically designed for network traffic analysis, incorporating temporal context through our Temporal-Enhanced Host-Connection Graph representation.
2. **Comprehensive Evaluation Framework**: The authors evaluate their approach across three diverse datasets (CIC IDS 2017, CIC DDoS 2019, and BCCC CPacket Cloud DDoS 2024), demonstrating robust performance across different attack types. While the first two datasets cover curated enterprise traffic, the third dataset provides more realistic network traffic where the approach still maintains strong performance.
3. **Practical Integration Strategy**: The authors provide detailed guidance for integrating GNN-based detection into existing data center infrastructure with minimal operational disruption.
4. **Performance Analysis**: The authors achieved F1 scores exceeding 95% across all evaluated datasets, significantly outperforming baseline approaches.

## 2. Related Work

### 2.1 Traditional DDoS Detection Methods

Traditional DDoS detection algorithms predominantly relied on signature-based and threshold-based approaches. Signature-based intrusion detection systems like SNORT (Li et al. 2019) have been widely deployed, which use predefined attack patterns to identify malicious traffic. These systems, while effective against known attacks, suffer from inability to detect novel attack variants and require constant signature updates. Statistical methods have also been employed, with researchers utilizing Hurst coefficients, autoregression models, and variance analysis to distinguish normal traffic from attack patterns (Hajtmanek et al. 2022). Fundamentally, these statistical techniques are threshold detectors at their core, they monitor a derived statistic and trigger only when it exceeds (or falls below) a limit. This design struggles with non-stationary baselines and adaptive adversaries.

Machine learning techniques emerged as a significant advancement over signature-based systems. Traditional ML approaches including Support Vector Machines, Random Forests, and k-nearest neighbors have demonstrated effectiveness in intrusion detection tasks. SVM-based approaches, often combined with optimization algorithms like Particle Swarm Optimization (Salam 2021), have shown particular promise in Internet of Things (IoT) DDoS detection scenarios. However, these methods typically operate on engineered features extracted from individual flows, failing to capture the relational dependencies crucial for understanding coordinated attack patterns.

Despite their widespread adoption, traditional methods face fundamental limitations in modern network environments. While computationally efficient, these methods suffer from an inability to adapt to legitimate traffic variations and evolving attack patterns. Furthermore, the heterogeneous nature of network traffic, with diverse device types and communication protocols, presents scalability challenges for traditional detection approaches. These limitations have driven the research community toward more adaptive and context-aware detection mechanisms.

### 2.2 Deep Learning for DDoS Detection

Convolutional Neural Networks have been extensively applied to DDoS detection, particularly for analyzing packet-level features. CNN-based approaches transform network packets into matrix representations, treating them as 2D images for classification between normal and malicious traffic. Recurrent neural networks, particularly Long Short-Term Memory (LSTM) networks, have shown promise in capturing temporal dependencies in network traffic. LSTM-based approaches, sometimes combined with Bayesian methods or optimization algorithms like Bacterial Colony Optimization (BCO), have demonstrated improved detection accuracy for time-series network data (Li and Lu 2019; Alamer and Shadadi 2023). Standardized RNN architectures have also been employed for detecting and classifying various types of network intrusions (Muhuri et al. 2020). These temporal models can effectively capture attack progression patterns but struggle with the inherently graph-structured nature of network communications.

Recent developments have explored more sophisticated deep learning architectures for DDoS detection. Generative Adversarial Networks (GANs) have been employed to generate synthetic attack data for improving detection model training, particularly effective for detecting novel attack variants (Shroff et al. 2022). Deep Belief Networks utilizing Restricted Boltzmann Machines have shown effectiveness in intrusion detection with limited labeled data (Manimurugan et al. 2020). Auto-encoders have been applied to anomaly detection tasks, learning normal behavior patterns to identify deviations (Nguimbous et al. 2019). Despite these advances, traditional deep learning approaches treat network flows as independent entities, missing crucial structural relationships between communicating hosts.

## 2.3 Graph-Based Methods

Graph Neural Networks represent a paradigm shift in DDoS detection by explicitly modeling the relational structure of network communications. Works by Zhou et al. (2020) and subsequent improvements using GCN and XG-BoT have shown significant advancement in botnet detection through topological pattern recognition. These approaches leverage the inherent graph structure of network communications to identify coordinated attack patterns that traditional methods might miss.

Advanced GNN architectures have been specifically developed for network intrusion detection scenarios. E-GraphSAGE (Lo et al. 2022) emerged as a notable advancement, enabling edge-level classification for flow-based detection. Custom Message Passing Neural Networks (MPNNs) have been designed with distinct aggregation functions for different node types, effectively handling heterogeneous network graphs (Pujol-Perich et al. 2022). Heterogeneous graph attention networks utilizing hand-designed meta-paths have demonstrated effectiveness in capturing complex attack semantics (Zhao et al. 2020). Self-supervised learning methods like Anomal-E (Caville et al. 2022) were developed to address the scarcity of labeled attack data, using mutual information maximization for representation learning. These specialized architectures address the unique challenges posed by network security applications.

While previous GNN based Network Intrusion Detection System (NIDS) approaches demonstrated significant advantages in leveraging network topology and relational structure, they were not able to simultaneously capture both fine-grained local attack signatures and global coordination patterns across multiple scales of the network hierarchy. Heterogeneous Graph U-Nets fills this gap through its hierarchical encode-decoder architecture with pooling and unpooling operations which enables multi-scale feature learning that can detect both localized flow anomalies and distributed coordination patterns characteristic of sophisticated DDoS attacks. The empirical results also validate this design choice as the Heterogenous Graph U-Nets performs the best on the realistic BCCC-cPacket-Cloud-DDoS 2024 dataset when compared to the baselines, showing that hierarchical multi-scale processing is crucial for accurate DDoS detection in complex cloud environments.

# 3. Heterogeneous Graph U-Nets Architecture for Traffic Analysis

## 3.1 Network Traffic Graph Representation

This paper's approach introduces a novel graph-based representation for network traffic that captures both structural and temporal dynamics of network communications. Unlike traditional flow-based approaches that analyze connections in aggregate, we model traffic as a heterogeneous graph comprising two node types: hosts (network endpoints identified by IP addresses) and flows (individual communication sessions). This dual-node representation preserves critical relational information that would otherwise be lost in conventional traffic analysis methods. The foundation builds upon the host-connection graph concept from Pujol-Perich et al. (2021), which demonstrated the effectiveness of graph-structured data for capturing attack patterns.

The graph structure follows a directed tripartite pattern where each flow connects to its source and destination hosts through directed edges. Host nodes are initialized with uniform features to maintain anonymity and prevent IP-specific learning that would limit generalization. Flow nodes contain rich statistical features extracted from packet-level data, including packet sizes, inter-arrival times, duration, and protocol characteristics. This heterogeneous structure reflects the fundamental asymmetry in network communications, where hosts serve as persistent entities while flows represent transient interactions.

The key innovation lies in our temporal-enhanced design that introduces a sliding window mechanism for maintaining historical context. As the system processes flows sequentially, it maintains a dynamic memory of previously encountered flows, enabling detection of sophisticated attacks that manifest across extended time periods. When constructing subgraphs for new flow batches, the system identifies and incorporates relevant historical flows that share common endpoints, creating temporal edges that link current and past network activities.

Memory management balances computational efficiency with temporal coverage through a configurable limit on historical flow connections. This approach prevents unbounded memory growth while ensuring sufficient context for detecting multi-stage attacks, slow-rate denial-of-service attempts, and coordinated botnet activities. The sliding window mechanism transforms static graph analysis into a dynamic process capable of capturing the evolutionary nature of modern cyberattacks.

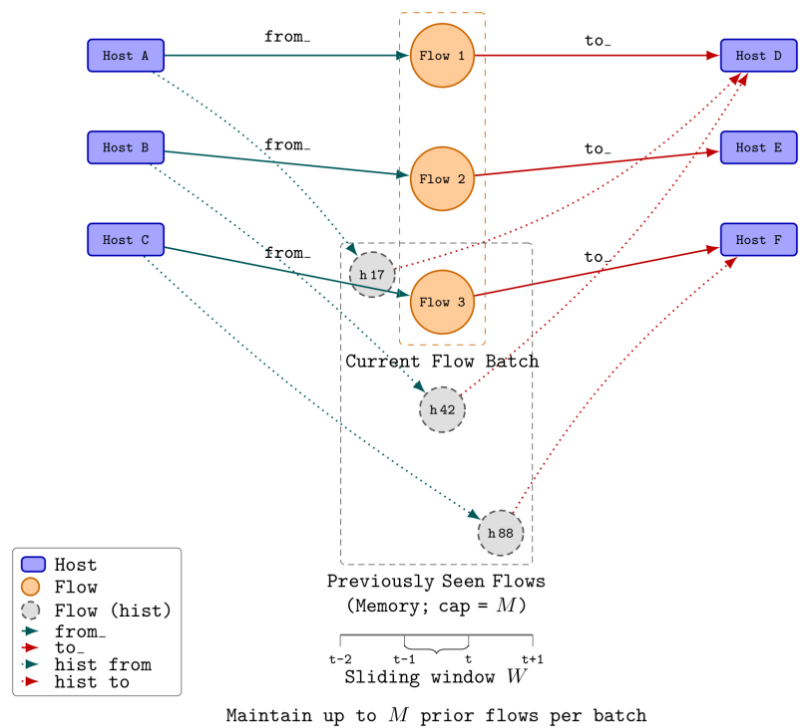

*Figure 1*: Network Traffic Graph

## 3.2 Heterogeneous Graph U-Nets Design

The Graph U-Nets (Gao and Shuiwang 2019) architecture adapts classical U-Net (Ronneberger et al. 2015) principles to heterogeneous graph processing through an encoder-decoder structure with skip connections. The encoder path implements hierarchical feature extraction using specialized graph convolution layers that respect the heterogeneous nature of the input while learning increasingly abstract representations. Initial embedding layers transform raw features into a unified hidden space using separate networks for host and flow nodes, followed by residual heterogeneous graph attention convolution layers that aggregate neighborhood information while maintaining distinct processing paths for different edge types.

A critical component is the heterogeneous attention pooling mechanism, which implements adaptive graph coarsening through attention-based scoring. Operating independently on host and flow nodes, it computes multi-head attention scores to identify the most informative nodes at each hierarchy level. Pooling ratios decay exponentially with depth (0.5 → 0.4 → 0.32), implementing progressive refinement that preserves more information in deeper layers. During training, controlled noise injection prevents deterministic patterns, while the multi-head design captures diverse importance criteria for robust node selection.

The bottleneck layer employs Graph Attention Networks (GAT) (Veličković et al. 2018) with multiple attention heads for enhanced message passing at the most abstract graph representation. This layer synthesizes global patterns before reconstruction begins, capturing both attack-specific structural patterns and normal network behavior baselines. The decoder path then implements progressive reconstruction by combining coarse-grained patterns from deeper layers with fine-grained details preserved through skip connections. The unpooling mechanism maps features back to original node positions using stored indices, while skip connections provide direct pathways for detailed information to bypass the bottleneck.

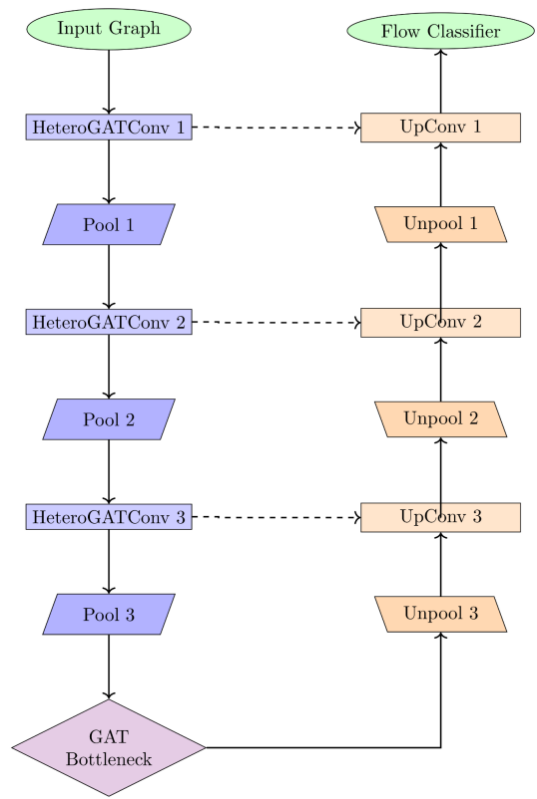

*Figure 2: The architecture of Heterogeneous Graph U-Nets*

The architecture concludes with a sophisticated three-layer classification head operating exclusively on flow nodes, implementing gradual dimensional reduction to binary decisions (benign/attack). Heavy regularization through layer normalization and dropout prevents overfitting and ensures robust generalization. This design is resilient against adversarial attacks by learning structural patterns rather than flow-level features, which theoretically should make it more resistant to evasion techniques that manipulate packet-level characteristics While we have not explicitly tested against adversarial attacks designed specifically for GNNs, the hierarchical nature of the architecture with multiple pooling levels should provide some inherent robustness against structural perturbations. The implementation is open source and publicly available at https://github.com/kartikeyas00/heterogeneous-graph-unets-ddos

# 4. System Design and Implementation

## 4.1 Data Processing Pipeline

The detection platform operates as a continuous, three-stage pipeline (Figure 3).

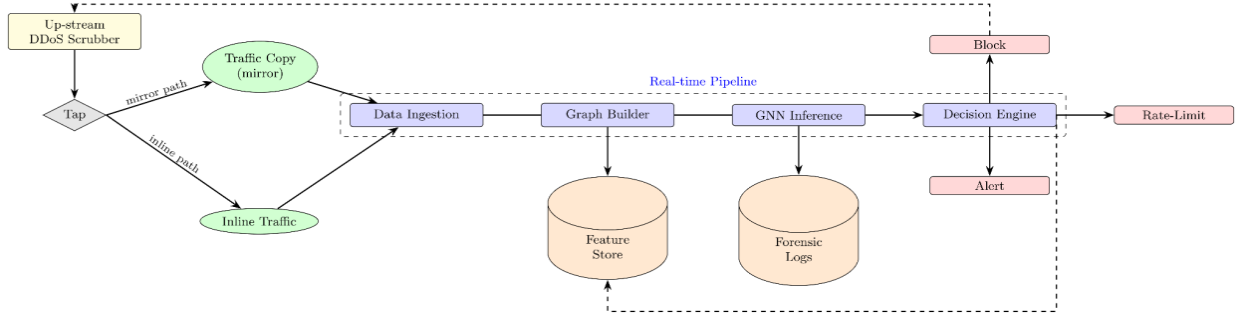

*Figure 3: System Design*

**Data Ingestion** occurs at the network perimeter, where lightweight collectors tap NetFlow, sFlow, IPFIX, or mirrored packets from switch SPAN ports. Each collector normalizes field names, verifies basic integrity, and suppresses duplicates before forwarding the cleansed records to a site-local buffering layer built on a distributed message-queue abstraction. This design decouples traffic spikes from downstream analytics and sustains multi-gigabit rates per node while keeping end-to-end latency in the sub-millisecond range.

In the **Graph Builder** stage, records are grouped into windows of size $\Delta t$ seconds or $B$ flows, whichever limit is reached first. Using the graph schema introduced in Section 3, the builder enriches every node with both static and dynamic traffic descriptors e.g., ACK-Flag Count, Initial-Window-Bytes-Forward, Minimum-Segment-Size-Forward, Forward-Inter-Arrival-Time {Mean, Max}, Flow-IAT {Mean, Max}, Forward-Packet-Length-Std, and specialised Web-DDoS indicators. Host attributes that remain stable across windows are cached in a key-value store to avoid recomputation, and a sliding-window mechanism re-uses previously emitted host nodes, thereby reducing memory churn. Feature tensors are concurrently streamed to an immutable object store so that any experiment can be reproduced exactly.

Batched graphs advance to the **GNN Inference** state where the engine runs the *HeteroGraph U-Net* model on dedicated accelerators. The engine returns a probability score $p \in [0, 1]$ for each flow; scores and attendant mis-classifications are archived in a forensic log that later seeds nightly retraining and drift-analysis jobs.

## 4.2 Integration with Data Center Infrastructure

The platform is designed to slip into a data-center network with minimal fabric changes and can operate **alone** or **in tandem with an upstream DDoS-mitigation service** (e.g., Akamai, Cloudflare, or an on-prem appliance). Two deployment modes are possible:

- **Inline mode.** A lightweight enforcement module, for example, an eBPF filter inside the software switch receives all traffic that survives the upstream scrubber and can drop or rate-limit flows immediately when the model assigns a probability above a configurable threshold $\tau\_auto$. Flows whose score falls into the grey zone $\tau\_analyst \leq p < \tau\_auto$ are tagged and mirrored to a security-operations console for human adjudication. Analyst feedback is fed back into the feature store so that future retraining cycles can tighten the grey zone.

- **Passive mode.** A hardware tap or SPAN port can clone packets or alternatively routers/switches can export flow telemetry (NetFlow/IPFIX/sFlow) to our collectors. The production packets stay on their normal path, so the model inspects them with zero added delay and zero chance of accidental drops. In this setup, the model's verdict is an **auxiliary signal** that analysts can compare with the primary DDoS provider's logs; any pattern the model flags repeatedly with high confidence can then be turned into a new blocking rule on the main DDoS platform.

Because each pipeline stage is stateless, horizontal scaling is a matter of adding additional processing nodes; a lightweight coordinator ensures that, should any node fail, its share of the workload is redistributed instantly, preventing ingestion lag. A staged roll-out is recommended:

1. **Phase 0 (alert-only).** Deploy on the management network and send detections to the Security Operations Center (SOC) without enforcement.
2. **Phase 1 (rate-limit).** Extend coverage to inter-data-center links; enable automated rate-limiting for flows with $p \geq \tau\_auto$.
3. **Phase 2 (full-block).** After a two-week user acceptance test confirms an acceptably low false-positive rate; permit hard blocking on tenant-edge routers; and feed recurring signatures back to the upstream DDoS provider for pre-emptive filtering.

In all phases the model's verdict is presented to SOC analysts as an extra datapoint rather than a replacement for the existing defense stack, allowing gradual trust-building and continuous improvement of both systems.

### 4.3 Mitigation and Response Mechanisms

The decision engine maps the model's confidence score to a tiered set of actions.

- **High-confidence detections** trigger automatic blocking at enforcement points such as edge firewalls, load balancers or the upstream DDoS-mitigation service by pushing short-lived policy updates via API/SDN. These controls auto-expire unless renewed by continuing evidence.
- **Medium-confidence events** invoke adaptive rate-limiting: a traffic-shaping policy is applied to the offending flow and tightened progressively while suspicious activity persists.
- **Low-confidence scores** generate notifications to the SOC via the existing Security Information and Event Management (SIEM) platform and analyze workflow, allowing analysts to review the evidence before any hard enforcement occurs.

Every mitigation decision can be archived in tamper-evident object storage for a fixed retention period. A recurring analytics job would fold these outcomes back into the feature store and would recommend when the model or policy thresholds should be retrained or tuned.

### 4.4 Analyst Workflow and Feedback Loop

When flow enters the grey zone ($\tau\_analyst \leq p < \tau\_auto$), the SOC analyst receives an alert containing:

- Flow metadata (source/dest IPs, ports, protocol, timestamp)
- Model confidence score and top contributing features
- Visual graph representation highlighting the suspicious subgraph

The analyst reviews this information and takes one of three actions: approve (legitimate traffic), block (confirmed attack), or rate-limit (suspicious but uncertain). These decisions are logged along with rationale that can be incorporated into nightly retraining and can progressively improve the model's accuracy on site-specific traffic patterns and reduce the grey zone boundaries over time.

# 5. Experimental Evaluation

## 5.1 Experimental Setup

### 5.1.1 Datasets

We use three publicly-available datasets, each captured under different operating conditions, to train and evaluate the detection model.

**CIC IDS 2017** (Sharafaldin et al. 2018)**:** After stripping every non-DDoS record, our working copy holds **2,399,345 flows**: 128,025 LOIC-generated attack flows and 2,271,320 benign flows. Each flow is described by the original 84 CICFlowMeter features—packet-length statistics, directional inter-arrival times, and flow duration among the most salient dimensions. Removing brute-force, Heartbleed (CVE-2014-0160), and web attacks yields a clean binary label space while preserving the realistic background traffic captured in the five-day experiment. This focused version allows the model to learn volumetric-flood behavior without distraction from unrelated threats.

**CIC-DDoS 2019** (Sharafaldin et al. 2019)**:** CIC-DDoS 2019 was built in a dual-day testbed that replayed 11 high-volume and low-volume attack families (e.g., DNS, LDAP, UDP-Lag, SYN) against a mix of Windows and Linux victims. All benign background traffic was kept, yielding a *fully DDoS-centric* corpus totalling **20,640,509 flows** across the 12 labelled classes, including the benign baseline. Its scale and diversity make it ideal for learning both reflection-based and exploitation-based flooding patterns in modern networks.

**BCCC-cPacket-Cloud-DDoS 2024** (Shafi et al. 2024): This dataset was captured inside an AWS VPC with VXLAN mirroring and parsed by NTLFlowLyzer, this modern corpus supplies **700,774 flows**: 228,469 DDoS, 59,106 "suspicious" and 413,199 benign. Its 322-dimensional feature set emphasises traffic-volume and window-size statistics, for example, packets_count, fwd_packets_count, payload_bytes_std, fwd_init_win_bytes and ack_flag_counts. Seventeen TCP-SYN variants (Valid-SYN, Flag-MIX, Killer-TCP, etc.) plus control/killer sequences provide a diverse attack portfolio, while benign days include e-mail, SSH/FTP and multimedia browsing. The authors retain all three labels to test the detector's ability to distinguish outright floods from ambiguous "suspicious" activity in elastic cloud environments.

### 5.1.2 Baseline Models

To compare the performance of Heterogeneous Graph U-Nets, we contrast it with two GNN-based intrusion detection models.

**E-GraphSAGE:** This baseline builds on inductive GraphSAGE but preserves the original flow-graph structure, treating every net-flow as an edge and every (IP, port) endpoint as a node, so the detector learns to classify edges directly from raw traffic features. The public implementation makes deployment straightforward and keeps the computational footprint low, which is useful when benchmarking against heavier graph models. Across the two evaluation corpora, E-GraphSAGE delivered strong results (weighted F1 = $0.942$ on CTU-13 and $0.978$ on ToN-IoT).

**GNN-RNIDS:** This model works on a *host-connection graph* that contains two node types: hosts and flows, linked with directed edges to preserve upstream and downstream semantics. A custom message-passing routine handles the heterogeneous neighbourhoods, allowing embeddings to capture structural signatures of multi-flow threats such as DDoS. On CIC-IDS 2017 the authors report a weighted F1 close to $0.99$, and the detector maintains its accuracy even when attackers perturb packet sizes or inter-arrival times which is an evidence of strong adversarial robustness.

### 5.1.3 Evaluation Metrics

We assess model performance with four standard classification metrics: accuracy, precision, recall, and F1-score. Accuracy captures overall correctness, precision gauges the proportion of predicted attacks that are truly malicious, and recall measures the fraction of real attacks our model successfully flags. F1-Score, which is the harmonic mean of precision and recall, balances false positives and false negatives to provide a single, robust indicator of effectiveness.

## 5.2 Results

The model performs 10-fold cross-validation: for each fold one-tenth of the data is held out as the test set, the remaining nine-tenths are shuffled, then 20 % of that training portion is stratified into a validation set, yielding roughly 72 % training, 18 % validation, and 10 % test per fold.

| TABLE 1: RESULTS ON CIC IDS 2017 | | | | |
|---|---|---|---|---|
| **Model** | **Accuracy** | **Precision** | **Recall** | **F1 Score** |
| *Het. Graph U-Nets* | *1.000 ± 0.000* | *1.000 ± 0.000* | *0.999 ± 0.002* | *0.999 ± 0.001* |
| GNN RNIDS | 0.999 ± 0.000 | 0.993 ± 0.006 | 0.999 ± 0.002 | 0.999 ± 0.001 |
| E-GraphSage | 1.000 ± 0.000 | 1.000 ± 0.000 | 1.000 ± 0.000 | 0.999 ± 0.000 |

Across the two *traditional* benchmarks **CIC IDS 2017** (TABLE 1) and **CIC DDoS 2019** (TABLE 2), all three models achieve exceptional performance, with every evaluation metric above 0.999 with tiny dispersion (standard deviation (SD)≤0.002). E-GraphSAGE posts the single highest recall on both corpora (0.9999 and 0.9998), indicating it rarely misses an attack once the traffic patterns are well-defined. The overlap of the ± SD intervals shows that the small gaps between models (e.g., E-GraphSAGE's slightly higher recall on CIC IDS 2017) are not statistically significant. In practice, every model is indistinguishable within measurement noise on these two well-curated datasets.

| TABLE 2: RESULTS ON CIC DDoS 2019 | | | | |
|---|---|---|---|---|
| **Model** | **Accuracy** | **Precision** | **Recall** | **F1 Score** |
| *Het. Graph U-Nets* | *1.000 ± 0.000* | *1.000 ± 0.000* | *1.000 ± 0.000* | *1.000 ± 0.000* |
| GNN RNIDS | 1.000 ± 0.000 | 1.000 ± 0.000 | 1.000 ± 0.000 | 1.000 ± 0.000 |
| E-GraphSage | 1.000 ± 0.000 | 1.000 ± 0.000 | 1.000 ± 0.000 | 1.000 ± 0.000 |

The *modern cloud* scenario (**BCCC CPacket Cloud DDoS 2024**) is markedly tougher. The background traffic is more diverse and attack signatures are subtler. Heterogeneous Graph U-Net achieves the best F1 score (0.960 ± 0.002), thanks to the highest precision (0.985 ± 0.002) while keeping recall above 0.945 ± 0.010. E-GraphSAGE sees its F1 tumble to (0.914 ± 0.021), primarily due to a precision drop (0.922 ± 0.029). GNN-RNIDS trails Heterogeneous Graph U-Net with a F1 score (0.940 ± 0.007), 0.015 below Heterogeneous Graph U-Net.

| TABLE 3: RESULTS ON BCCC-cPacket-Cloud-DDoS 2024 | | | | |
|---|---|---|---|---|
| Model | Accuracy | Precision | Recall | F1 Score |
| *Het. Graph U-Net* | *0.972 ± 0.001* | *0.985 ± 0.002* | *0.945 ± 0.010* | *0.960 ± 0.002* |
| GNN RNIDS | 0.957 ± 0.005 | 0.946 ± 0.016 | 0.982 ± 0.018 | 0.940 ± 0.007 |
| E-GraphSage | 0.947 ± 0.014 | 0.922 ± 0.029 | 0.994 ± 0.010 | 0.914 ± 0.021 |

These findings show that when attack patterns are clean and voluminous, lightweight inductive models like E-GraphSAGE suffice, but once traffic grows multifaceted and irregular (as in cloud environments), the explicit host-flow heterogeneity and hierarchical pooling of Heterogeneous Graph U-Net materially reduces false positives without sacrificing recall. The consistent precision and recall along with minimal standard deviation (< 0.005) on the cloud dataset showcases Heterogeneous Graph U-Net's ability for practical real-world deployment where stability is most important.

## 6. Limitations and Future Work

### 6.1 Current Limitations

Despite strong performance, several limitations warrant acknowledgment:

- **Open-source Data Constraints**: Training on publicly available datasets may not fully capture specific data center attack patterns.
- **Scalability Boundaries**: While efficient, processing graphs for networks exceeding 100,000 concurrent flows requires distributed implementations,which we have not yet developed or tested.
- **Feature Engineering**: Optimal feature selection remains dataset-dependent, requiring domain expertise.
- **Adversarial Robustness**: While the architecture should theoretically provide resilience against adversarial attacks through its hierarchical structure, we have not explicitly tested against GNN-specific adversarial examples or structural perturbations designed to evade detection.
- **Performance Benchmarking**: While the system architecture is designed for low-latency operation, we did not measure actual inference times or throughput rates in our experiments. Production deployment would require comprehensive performance benchmarking to validate the system's ability to process traffic efficiently.
- **Ablation Study**: We have not conducted an ablation study to determine which components of the Graph U-Nets architecture (hierarchical pooling, skip connections, heterogeneous processing, or temporal sliding window) contribute most to the performance gains. Understanding the relative importance of each component would help optimize the architecture and potentially reduce computational overhead.

### 6.2 Future Research Directions

Several promising avenues for future work are:

- **Adaptive Learning**: Incorporate online or continual-learning routines so the detector can update itself as new attack behaviors emerge.
- **Multi-site Coordination**: Explore cross-site or federated learning schemes that allow multiple datacenters to share model updates without exposing raw traffic. This would require addressing significant privacy challenges including preventing model inversion attacks, ensuring differential privacy in gradient updates, and coordinating update synchronization.

- **Hardware Acceleration**: Evaluate Graphics Processing Units (GPU) and Field-Programmable Gate Arrays (FPGA) pipelines to push inference latency into the sub-millisecond range for high-throughput networks.
- **Encrypted Traffic Analysis:** As traffic encryption becomes ubiquitous, future work should explore how to adapt the graph-based approach when traditional flow features are obscured. This might involve leveraging timing-based side channels, TLS handshake patterns, and graph structural features that persist despite encryption. New statistical features based on encrypted payload characteristics would need to be developed and validated.
- **Zero-Day Attack Evaluation**: Conduct systematic evaluation of the model's performance against completely novel attack types through leave-one-attack-out cross-validation and testing on emerging threats not represented in current datasets. Depending on dataset availability, extending this analysis to known DDoS attacks not seen within the aforementioned datasets e.g. Protocol Exploits (e.g., CoAP, WS-DD, ARMS, Jenkins) - 2020, TP240PhoneHome - 2023, HTTP/2 Rapid Reset - 2023, etc.
- **Ablation Study**: Conduct systematic ablation experiments to quantify the contribution of each architectural component (hierarchical pooling layers, skip connections, heterogeneous node processing, temporal sliding window) to overall performance. This would guide architectural optimization and potentially enable simpler, more efficient variants.
- **Adversarial Robustness Testing**: Evaluate the model against adversarial attacks specifically designed for GNNs, including graph structure perturbations, node feature poisoning, and edge manipulation attacks.

# 7. Conclusion

This paper presented a novel Graph U-Nets architecture for DDoS detection in datacenter environments. Our study shows that modelling network traffic as a heterogeneous host-flow graph and processing it with a Graph U-Net yields exceptional detection accuracy across both legacy and modern DDoS corpora. On the well-curated CIC IDS 2017 and CIC DDoS 2019 benchmarks, all contenders already operate near perfection, yet the model matches this performance while keeping the harmonic balance of precision and recall tight, demonstrating that the richer representation does not sacrifice efficiency in straightforward scenarios. The Graph U-Nets algorithm excels in the cloud-native BCCC CPacket dataset where the Graph U-Nets' hierarchical pooling and type-aware message passing cut false-positives dramatically, lifting precision to 0.985 and F1-score to 0.960, improving over the next-best baseline (GNN RNIDS) by 3.9 and 2.0 points respectively while trading some recall for a substantially lower false-positive rate.

Beyond raw metrics, the proposed pipeline integrates into existing data-center operations; it scales horizontally, supports "alert-only" or inline enforcement, and preserves sub-millisecond latency through decoupled ingestion and batching. This flexibility allows data center operators to start conservatively and progress toward automated blocking as confidence grows. Importantly the model's superior precision in irregular, multi-tenant traffic means fewer distracting alerts for analysts and quicker mitigation of real threats, translating into tangible availability and reputational benefits.

Nevertheless, several avenues remain open. Our evaluation still relies on public datasets whose attack mix may differ from proprietary environments; future work should explore online or federated learning to adapt to unseen patterns. We also aim to profile throughput on very large graphs (>100 k concurrent flows) and investigate hardware acceleration. Finally, enriching the feature space with encrypted-traffic markers and integrating cross-site threat intelligence could further harden defenses as adversaries evolve.

# References

Alamer, Latifa, and Ebtesam Shadadi. 2023. "DDoS Attack Detection Using Long-Short Term Memory with Bacterial Colony Optimization on IoT Environment." *Journal of Internet Services and Information Security* 13, no. 1: 44–53.

Caville, Evan, Wai Weng Lo, Siamak Layeghy, and Marius Portmann. 2022. "Anomal-E: A Self-Supervised Network Intrusion Detection System Based on Graph Neural Networks." *Knowledge-Based Systems* 258: 110030.

Data Center Knowledge. 2024. "DDoS Attacks: Data Centers Caught in the Crosshairs." *Data Center Knowledge.* February 20, 2024. https://www.datacenterknowledge.com/cybersecurity/ddos-attacks-data-centers-caught-in-the-crosshairs (accessed July 28, 2025).

Gao, Hongyang, and Shuiwang Ji. 2019. "Graph U-Nets." In *Proceedings of the 36th International Conference on Machine Learning (ICML)*, 2083–2092. PMLR.

Hajtmanek, Roman, Martin Kontšek, Juraj Smieško, and Jana Uramová. 2022. "One-Parameter Statistical Methods to Recognize DDoS Attacks." *Symmetry* 14, no. 11: 2388.

Kleyman, Bill. 2023. "Why You Need DDoS Protection in a Connected World." *Data Center Frontier.* January 27, 2023. https://www.datacenterfrontier.com/sponsored/article/21545872/a10-why-you-need-ddos-protection-in-a-connected-world (accessed July 28, 2025).

Li, Wenjuan, Steven Tug, Weizhi Meng, and Yu Wang. 2019. "Designing Collaborative Blockchained Signature-Based Intrusion Detection in IoT Environments." *Future Generation Computer Systems* 96: 481–489.

Li, Yan, and Yifei Lu. 2019. "LSTM-BA: DDoS Detection Approach Combining LSTM and Bayes." In *2019 Seventh International Conference on Advanced Cloud and Big Data (CBD)*, 180–185. IEEE.

Lo, Wai Weng, Siamak Layeghy, Mohanad Sarhan, Marcus Gallagher, and Marius Portmann. 2022. "E-GraphSAGE: A Graph Neural Network Based Intrusion Detection System for IoT." In *NOMS 2022—IEEE/IFIP Network Operations and Management Symposium*, 1–9. IEEE.

Manimurugan, S., Saad Al-Mutairi, Majed Mohammed Aborokbah, Naveen Chilamkurti, Subramaniam Ganesan, and Rizwan Patan. 2020. "Effective Attack Detection in Internet of Medical Things Smart Environment Using a Deep Belief Neural Network." *IEEE Access* 8: 77396–77404.

Muhuri, Pramita Sree, Prosenjit Chatterjee, Xiaohong Yuan, Kaushik Roy, and Albert Esterline. 2020. "Using a Long Short-Term Memory Recurrent Neural Network (LSTM-RNN) to Classify Network Attacks." *Information* 11, no. 5: 243.

Nguimbous, Yves Nsoga, Riadh Ksantini, and Adel Bouhoula. 2019. "Anomaly-Based Intrusion Detection Using Auto-Encoder." In *2019 International Conference on Software, Telecommunications and Computer Networks (SoftCOM)*, 1–5. IEEE.

Pujol-Perich, David, José Suárez-Varela, Albert Cabellos-Aparicio, and Pere Barlet-Ros. 2022. "Unveiling the Potential of Graph Neural Networks for Robust Intrusion Detection." *ACM SIGMETRICS Performance Evaluation Review* 49, no. 4: 111–117.

Ronneberger, Olaf, Philipp Fischer, and Thomas Brox. 2015. "U-Net: Convolutional Networks for Biomedical Image Segmentation." In *Medical Image Computing and Computer-Assisted Intervention – MICCAI 2015*, 234–241. Cham: Springer International Publishing.

Salam, Mahmoud A. 2021. "Intelligent System for IoT Botnet Detection Using SVM and PSO Optimization." *J. Intell. Syst. Internet Things* 3, no. 2: 68–84.

Shafi, MohammadMoein, Arash Habibi Lashkari, Vicente Rodriguez, and Ron Nevo. "Toward generating a new cloud-based Distributed Denial of Service (DDoS) dataset and cloud intrusion traffic characterization." *Information* 15, no. 4 (2024): 195.

Sharafaldin, Iman, Arash Habibi Lashkari, and Ali A. Ghorbani. "Toward generating a new intrusion detection dataset and intrusion traffic characterization." *ICISSp* 1, no. 2018 (2018): 108-116.

Sharafaldin, Iman, Arash Habibi Lashkari, Saqib Hakak, and Ali A. Ghorbani. "Developing realistic distributed denial of service (DDoS) attack dataset and taxonomy." In *2019 international carnahan conference on security technology (ICCST)*, pp. 1-8. IEEE, 2019.

Shroff, Jugal, Rahee Walambe, Sunil Kumar Singh, and Ketan Kotecha. 2022. "Enhanced Security Against Volumetric DDoS Attacks Using Adversarial Machine Learning." *Wireless Communications and Mobile Computing* 2022, no. 1: 5757164.

Veličković, Petar, Guillem Cucurull, Arantxa Casanova, Adriana Romero, Pietro Liò, and Yoshua Bengio. 2018. "Graph Attention Networks." In *International Conference on Learning Representations (ICLR)*.

Volico Data Centers. 2024. "DDoS Protection Trends Among Data Center and Colocation Providers." *Volico Data Centers.* March 26, 2024. https://www.volico.com/ddos-protection-trends-among-data-center-and-colocation-providers/(accessed July 28, 2025).

Zhao, Jun, Xudong Liu, Qiben Yan, Bo Li, Minglai Shao, and Hao Peng. 2020. "Multi-Attributed Heterogeneous Graph Convolutional Network for Bot Detection." *Information Sciences* 537: 380–393.

Zhou, Jiawei, Zhiying Xu, Alexander M. Rush, and Minlan Yu. 2020. "Automating Botnet Detection with Graph Neural Networks." *arXiv* preprint arXiv:2003.06344.

Ahmed, Aseem and Bella, Abdeslam. 2024. "Akamai's Behavioral DDoS Engine: A Breakthrough in Modern DDoS Mitigation." *Akamai.* November 07, 2024. https://www.akamai.com/blog/security/akamais-behavioral-ddos-engine-breakthrough-in-modern-ddos-mitigation (accessed July 28, 2025).